\documentstyle[12pt,fullpage]{article}
\def\be{\begin{equation}}
\def\ee{\end{equation}}
\def\beq{\begin{eqnarray}}
\def\eeq{\end{eqnarray}}
\def\lsim{\:\raisebox{-0.5ex}{$\stackrel{\textstyle<}{\sim}$}\:}
\def\gsim{\:\raisebox{-0.5ex}{$\stackrel{\textstyle>}{\sim}$}\:} 
\begin{document}
\begin{flushright}
TIFR/TH/96-58\\
hep-ph/yymmnn
\end{flushright}
\bigskip
\begin{center}
{\Large{\bf Constraints on the Charged Higgs Sector from the Tevatron
Collider Data on Top Quark Decay}} \\[2cm]
{\large Monoranjan Guchait\footnote{e-mail:guchait@@theory.tifr.res.in} 
and D.P. Roy\footnote{e-mail:dproy@@theory.tifr.res.in}} \\[1cm]
Theoretical Physics Group \\
Tata Institute of Fundamental Research \\
Homi Bhabha Road, Mumbai 400 005, India
\end{center}
\bigskip\bigskip
\begin{center}
\underbar{\bf Abstract}
\end{center}
\medskip

The top quark data in the lepton plus $\tau$ channel offers a viable
probe for the charged Higgs boson signal.  We analyse the recent
Tevatron collider data in this channel to obtain a significant limit
on the $H^\pm$ mass in the large $\tan\beta$ region.\\

\vspace{5cm}

Pacs Nos: 11.30.Pb,13.35.Dx,14.65.Ha

\newpage

The discovery of the top quark signal at the Tevatron collider [1,2]
has generated a good deal of current interest in the search of new
particles in top quark decay.  The large mass of top offers the
possibility of carrying on this search to a hitherto unexplored mass
range for these particles.  In particular the top quark decay is known
to provide by far the best discovery limit for one such new particle,
i.e. the charged Higgs boson [3] of the minimal supersymmetric
standard model (MSSM).  The signature of the charged Higgs boson in
top quark decay is based on its preferential coupling to the $\tau$
channel in contrast to the universal $W$ boson coupling.  Thus a
departure from the universality prediction can be used to separate the
charged Higgs boson signal from the $W$ background in 
\be
t \rightarrow bH (bW) \rightarrow b\tau\nu.
\ee
In particular, the top quark decay into the $\tau$ lepton channel
provides a promising signature for charged Higgs boson in the region
\be
\tan\beta \gsim m_t/m_b,
\ee
where $\tan\beta$ denotes the ratio of the vacuum expectation values
of the two Higgs doublets in MSSM.

In this note we shall analyse the recent CDF data on $t\bar t$ decay
events in the $\ell\tau$ channel, where $\ell$ denotes $e$ and $\mu$
[2,4,5].  This channel has the advantage of a low background.  As
we shall see below, the number of $t\bar t$ events in this dilepton
channel relative to the $\ell \ +$ multijet channel gives a significant
lower bound on the $H^\pm$ mass in the large $\tan\beta$ region (2).

In the diagonal KM matrix approximation, the charged Higgs boson
couplings to the fermions are given by
\be
{\cal L} = {g \over \sqrt{2} m_W} H^+ \left\{\cot\beta m_{ui} \bar u_i d_{iL}
+ \tan\beta m_{di} \bar u_i d_{iR} + \tan\beta m_{\ell i} \bar\nu_i
\ell_{iR} \right\} + {\rm h.c.}
\ee
where the subscript $i$ denotes quark and lepton generation.  The
leading log QCD correction is taken into account by substituting the
quark mass parameters by their running masses evaluated at the $H^\pm$
mass scale [6].  The resulting decay widths are
\beq
\Gamma_{t \rightarrow bW} &=& {g^2 \over 64\pi m^2_W m_t}
\lambda^{1\over2} \left(1,{m^2_b \over m^2_t}, {m^2_W \over
m^2_t}\right) \nonumber \\[2mm]
& & \left[m^2_W (m^2_t + m^2_b) + (m^2_t - m^2_b)^2 - 2m^4_W\right]
\eeq
\beq
\Gamma_{t \rightarrow bH} &=& {g^2 \over 64\pi m^2_W m_t}
\lambda^{1\over2} \left(1,{m^2_b \over m^2_t},{m^2_H \over
m^2_t}\right) \nonumber \\[2mm]
& & \left[(m^2_t \cot^2 \beta + m^2_b \tan^2\beta) (m^2_t + m^2_b -
m^2_H) - 4m^2_t m^2_b\right]
\eeq
\be
\Gamma_{H \rightarrow \tau\nu} = {g^2 m_H \over 32 \pi m^2_W} m^2_\tau
\tan^2\beta 
\ee
\be
\Gamma_{H \rightarrow c\bar s} = {3g^2 m_H \over 32\pi m^2_W} (m^2_c
\cot^2\beta + m^2_s \tan^2\beta).
\ee
They clearly show a large branching fraction for $t \rightarrow bH$
decay at $\tan\beta \lsim 1$ and $\tan\beta \gsim m_t/m_b$, while the
branching fraction for $H \rightarrow \tau\nu$ decay is $\simeq 1$ at
$\tan\beta \gg 1$.  Thus one expects a large charged Higgs boson
signal in top quark decay into the $\tau$ channel (1) in the large
$\tan\beta$ region (2).

In the present analysis, we shall concentrate in the $\tan\beta \gg 1$
region, for which the charged Higgs boson decays dominantly into
$\tau\nu$.  The basic process of interest is $t\bar t$ production via
quark-antiquark (or gluon-gluon) fusion, followed by their decays into
charged Higgs or $W$ boson channels, i.e.
\be
q\bar q \rightarrow t\bar t \rightarrow b\bar b (W^+ W^-, W^\pm H^\mp,
H^+ H^-).
\ee
The most important $t\bar t$ signal is observed in the $\ell \ +$
multijet channel [1,2].  It comes from the $W^+W^-$ final state with a
branching fraction of 24/81.  The corresponding branching fraction
from this final state into the $\ell\tau$ channel is only 4/81.
However there would be an additional contribution to the $\ell\tau$
channel from the $W^\pm H^\mp$ final state with a large branching
fraction of 4/9, which is to be weighted of course by the ratio
$\Gamma_{t \rightarrow bH}/\Gamma_{t \rightarrow bW}$.  Thus a comparison
of the number of $t\bar t$ events in the two channels leads to an upper
limit on this ratio, which can be translated into lower limit on
$H^\pm$ mass for a given $\tan\beta$.

For a quantitative estimate of the above limit, we have to consider
the various kinematic cuts and detection efficiencies [2,5].  Our
analysis is based on a parton level Monte Carlo simulation of $t\bar
t$ production using the quark and gluon structure functions of [7].
This is followed by the decays
\be
t \ {\buildrel W^+ \over \longrightarrow} \ b\ell\nu, \ \bar t \ {\buildrel
W^- \over \longrightarrow} \ \bar b q \bar q'
\ee
and vice versa for the $\ell \ +$ multijet channel.  The quark jets are
merged according to the CDF jet cone algorithm of $\Delta R = (|\Delta
\eta|^2 + |\Delta \phi|^2)^{1/2} = 0.7$.  The resulting final state is
required to satisfy the CDF cuts [2]
\be
p^\ell_T > 20 \ {\rm GeV}, \ |\eta_\ell| < 1, \ {E\!\!\!/}_T > 20 \ {\rm
GeV} \ {\rm and} \ n_{\rm jet} \geq 3 \ ({\rm with} \ E^j_T > 15
\ {\rm GeV}, \ |\eta_j| < 2).
\ee
We estimate the efficiency factor for these kinematic and topological
cuts to be 52\% for $m_t = 175 \ {\rm GeV}$.  This has to be supplemented by
the following CDF efficiency factors [2,5,8];
\be
\epsilon^\ell_{tr} = .93, \ \epsilon^\ell_{id} = .87, \
\epsilon^\ell_{iso} = .9, \ \epsilon_b = .4, \ \epsilon_{az} = .85,
\ee
corresponding to lepton triggering, identification and isolation-cut
along with those due to SVX $b$-tagging and azimuthal gaps in the
detector.  The combined efficiency factor is 12.8\%, in reasonable
agreement with the CDF estimate of 11.8\% [2] including hadronisation
and a more exact detector simulation.

The measured $\bar tt$ cross-section from this channel, including SLT
$b$-tagging, is [2]
\be
\sigma_t = 7.5 \pm 1.5 \ pb.
\ee
This is 40-50\% higher than the NLO QCD prediction for $m_t = 175$ GeV
[9].  We shall use this cross-section for normalisation.  Thus our
results will be independent of any theoretical model for the $\bar tt$
cross-section.  It will only depend on the preferencial $H^\pm$
coupling to the $\tau$ channel vis-a-vis the universal $W^\pm$ boson
coupling.  It should be noted here that the above cross-section
corresponds to the $WW$ contribution to the $\bar tt$ cross-section,
represented by the 1st term of (8), since any contribution from the
other terms would have very small detection efficiency for this
channel. 

The $\ell\tau$ channel of our interest corresponds to the decays
\be
t \ {\buildrel W^+ \over \longrightarrow} \ b\ell\nu, \ \bar t \
{\buildrel W^- (H^-) \over \longrightarrow} \ \bar b\tau\nu
\ee
and vice versa, where the $\tau$ lepton coming from $W(H)$ decay has a
definite polarization $P_\tau = -1 (+1)$.  It is identified in its
hadronic decay mode as a thin jet containing 1 or 3 charged prongs
[4,5].  This accounts for a $\tau$ branching fraction of about 64\%
[10].  The dominant contributions come from
\be
\tau \rightarrow \pi\nu (12\%), \ \rho\nu (25\%), \ a_1 \nu (15\%),
\ee
adding up to a little over 80\% of the hadronic $\tau$ decay.  We
shall combine these three decay modes and scale up their sum by 20\%
to simulate hadronic $\tau$-decay.  The decay distributions are simply
given by [11]
\be
{d\Gamma_\pi \over \Gamma_\pi d \cos\theta} = {1\over2} (1 + P_\tau
\cos\theta), 
\ee
\be
{d\Gamma_v \over \Gamma_v d \cos\theta} = {1\over2} \left(1 +
{m^2_\tau - 2m^2_v \over m^2_\tau + 2m^2_v} P_\tau
\cos\theta\right), \ v = \rho,a_1
\ee
where $\theta$ is the direction of the decay hadron in $\tau$ rest
frame relative to the $\tau$ line of flight.  It is related to the
fraction $x$ of $\tau$ momentum carried by the hadron,
\be
\cos\theta = {2x - 1 - m^2_\pi,v/m^2_\tau \over 1 - m^2_\pi,
v/m^2_\tau}.
\ee
This decay hadron momentum is refered to below as $p^\tau$.

The resulting final state is required to satisfy the CDF kinematic
cuts [4,5], 
\be
p^\ell_T > 20 \ {\rm GeV}, \ |\eta_\ell| < 1, \ p^\tau_T > 15 \ {\rm
GeV}, \ |\eta_\tau| < 1.2.
\ee
The corresponding efficiency factors are shown in the first column of
Table 1 for the $WW$ and $WH$ contributions with different charged
Higgs boson masses.  It includes the hadronic branching fraction of
$\tau$ along with a factor of 0.8 due to azimuthal gaps in
the detector, resulting in 15\% (5\%) loss to $\ell (\tau)$ detection
efficiency [8].  The opposite polarizations of $\tau$ coming from $W$
and $H$ decays results in a somewhat larger efficiency factor for the
latter, which increases further with increasing $H$ mass.  The second
column shows the CDF efficiency factors for the lepton trigger,
isolation and idendification as well as the $\tau$ identification.
These are expected to be essentially process independent.  The last
column shows the efficiency factors for the topological and
missing-$E_T$ cuts [4,5]
\be
n_{\rm jet} \geq 2 \ ({\rm with} \ E^j_T > 10 \ {\rm GeV}, \ |\eta_j|
< 2),
\ee
\be
H = p^\ell_T + p^\tau_T + {E\!\!\!/}_T + \sum_j E^j_T > 180 \ {\rm
GeV}, 
\ee
\be
\sigma_{E\!\!\!/_T} = E\!\!\!/_T/\sqrt{p^\ell_T + p^\tau_T + \sum_j
E^j_T} > 3 \ {\rm GeV}^{1/2}.
\ee
For the $WW$ contribution, the efficiency factors for the kinematic
and topological cuts from the CDF simulation [5] are shown in
parantheses for comparison.  For both cases they are 15-20\% below our
MC estimates, which indicate an overall error of $\sim 30$\% in our MC
result. 

The product of the efficiency factors in the three columns of Table 1
gives the overall acceptance factor for the $\ell\tau$ channel.  This
is to be multiplied by the branching fraction of 4/81 for the $WW$
contribution and 4/9 times $\Gamma_{t \rightarrow bH}/\Gamma_{t \rightarrow
bW}$ for the $WH$.  The resulting factor gives the corresponding
$\ell\tau$ cross-section as a fraction of the $\sigma_t$ of (12).

Fig. 1 shows the predicted cross-section in the $\ell\tau$ channel
against $\tan\beta$ for several charged Higgs boson masses.  The scale
on the right shows the corresponding number of events for the
accumulated CDF luminosity of 110 fb$^{-1}$.  The prediction includes
the $WW$ contribution of 14 fb, i.e. 1.5 events.  The corresponding
number from the CDF simulation is 1.2 events [5,8].  It may be noted
that the dominant contribution comes from $WW$ for $\tan\beta = 5 -
10$, where $t \rightarrow bH$ width has a pronounced dip.  However the
$WW$ is overwhelmed by the $WH$ contribution, when kinematically
allowed, for $\tan\beta \gsim m_t/m_b$.  The preliminary CDF data
shows 4 events in this channel against a background of $2 \pm 0.4$
[2,4,5].  The corresponding 95\% CL limit of 7.7 events [10,12] is
indicated in the figure.  This implies a $H^\pm$ mass limit of 100 GeV
for $\tan\beta \geq 40$, going up to 120 GeV for $\tan\beta \geq 50$.
One may scale down the predicted cross-section by 30\% to account for
the difference between the acceptance factors of CDF simulation and
ours.  This would correspond to an upward shift of the above
$\tan\beta$ limit by about 10 units for a given $H^\pm$ mass.
Nonetheless it would still represent a very significant constraint on
the charged Higgs boson mass in the large $\tan\beta$ region.

The mass limits of Fig. 1 for different values of $\tan\beta$ are
converted into a 95\% exclusion contour in the $m_H - \tan\beta$ plane
in Fig. 2.  As mentioned above, a 30\% reduction in the signal
cross-section would correspond to a rightward shift of this contour by
roughly 10 units in $\tan\beta$.  The scale on the right shows the
corresponding pseudo-scalar Higgs mass $m_A$ from the MSSM mass
relation, $m^2_{H^\pm} - m^2_{A} = m^2_{W}$,  
at the tree level [13].  The radiative correction to this
mass relation is known to be no more than a few GeV.  One sees from
this figure that a relatively light pseudo-scalar mass $(m_A \leq 60 \
{\rm GeV})$ is disallowed for $\tan\beta \geq 40$.  It is precisely in
this region of $m_A$ and $\tan\beta$ that one expects to get a
significant radiative correction to $R_b$ $(\Gamma^b_z/\Gamma^{\rm
had}_z)$ from the Higgs sector of MSSM [14].  Thus the so called large
$\tan\beta$ solution to the (so called!) $R_b$ anomaly seems to be
strongly disfavoured by the above CDF data.

Recently the CDF collaboration has obtained a limit on the charged
Higgs mass in the large $\tan\beta$ region [15] on the basis of their
analogous data in the inclusive $\tau$ channel.  Thus it is
instructive to compare the relative merits of the two channels for
probing the charged Higgs signal.  The inclusive $\tau$ channel
corresponds to a larger branching fraction than the $\ell\tau$ channel
analysed here.  However it is compensated by much stronger
experimental cuts, required to control the background.  Consequently
the final signal cross-section in the inclusive $\tau$ channel is
similar to that in the $\ell\tau$ channel.  This can be seen by
comparing the predicted cross-sections in the two channels in the
region of $\tan\beta = 5 - 10$.  The reason we get a much larger
signal cross-section in the large $\tan\beta$ region compared to [15]
and hence a stronger mass limit is due to the different normalisation
procedure followed in the two cases.  We use the $t\bar t$
cross-section in the $WW$ decay mode as measured via the lepton plus
$\geq 3$ jets channel for our normalisation, while the QCD prediction
for the inclusive $t\bar t$ cross-section is used for normalisation in
[15].  The former method is evidently more powerful in the large
$\tan\beta$ region and should be used in the analysis of the inclusive
$\tau$ channel as well.

It should be noted here that even with stronger cuts the number of
estimated background events in the inclusive $\tau$ channel are 5
times larger than in the $\ell\tau$ channel, for equal luminosity
[5,15].  Thus the $\ell\tau$ channel will be clearly more advantageous
at Tevatron upgrade, which promises a 20 times higher luminosity along
with a 2 times larger $t\bar t$ cross-section.  In particular the
$\ell\tau$ channel with $b$-tagging seems to be practically free from
non-top background [4,5].  The main background in this case is from
the $t\bar t$ decay via the $WW$ mode.  This can be suppressed
relative to the $H^\pm$ signal by exploiting the opposite
polarizations of $\tau$ lepton in the two cases [16].  Thus the
$\ell\tau$ channel with $b$-tagging is best suited for the charged
Higgs boson search at the Tevatron upgrade as well as the LHC.

In summary, the $t\bar t$ data in the $\ell\tau$ channel is well
suited to probe for a charged Higgs boson signal because of the small
background.  On the basis of the recent CDF data in this channel we
can already get a significant limit on the $H^\pm$ mass in the large
$\tan\beta$ region.  With a much higher luminosity expected at the
Tevatron upgrade the probe can be extended over a significantly wider
range of $H^\pm$ mass and $\tan\beta$.

It is a pleasure to thank Prof. M. Hohlmann of the CDF collaboration
and Prof. N.K. Mondal of D${\rm O}\!\!\!\!/$ for many helpful discussions.

\newpage

\begin{center}
\underbar{\bf References}
\end{center}
\bigskip
\begin{enumerate}
\item[{1.}] CDF Collaboration: F. Abe et. al., Phys. Rev. Lett. 74,
2626 (1995); \\
D${\rm O}\!\!\!\!/$ Collaboration: S. Abachi et. al.,
Phys. Rev. Lett. 74, 2632 (1995). 
\item[{2.}] P. Tipton, 28th Intl. Conf. on High Energy Physics, Warsaw
(1996). 
\item[{3.}] V. Barger and R.J.N. Phillips, Phys. Rev. D41, 884 (1990);
\\ A.C. Bawa, C.S. Kim and A.D. Martin, Z. Phys. C47, 75 (1990); \\
R.M. Godbole and D.P. Roy, Phys. Rev. D43, 3640 (1991); \\
R.M. Barnett, R. Cruz, J.F. Gunion and B. Hubbard, Phys. Rev. D47,
1048 (1993).
\item[{4.}] CDF Collaboration: S. Leone, XI Topical Workshop on $\bar
pp$ Collider Physics, Padova (1996), Fermilab-Conf. 96/195-E.
\item[{5.}] CDF Collaboration: M. Hohlmann, Lake Louise Winter School
(1996). 
\item[{6.}] M. Drees and D.P. Roy, Phys. Lett. B269, 155 (1991); \\
D.P. Roy, ibid B283, 403 (1992).
\item[{7.}] A.D. Martin, R.G. Roberts and W.J. Stirling,
Phy. Lett. B306, 145 (1993) and B309, 492 (1993).
\item[{8.}] M. Hohlmann, Private Communication.
\item[{9.}] E. Berger and H. Contoparagos, Phys. Lett. B361, 115
(1995); \\ S. Catani et. al., hep-ph/9602208 (1996).
\item[{10.}] Particle Data Group: R. M. Barnett et al, Phys. Rev. 
D54, 1 (1996). 
\item[{11.}] B.K. Bullock, K. Hagiwara and A.D. Martin,
Phys. Rev. Lett. 67, 3055 (1991); \\
D.P. Roy, Phys. Lett. B277, 183 (1992).
\item[{12.}] O. Helene, Nucl. Inst. and Meth., 212, 319 (1983).
\item[{13.}] J.F. Gunion, H.E. Haber, G. Kane and S. Dawson, The Higgs
Hunters' Guide, Addison Wesley, Reading, MA (1990).
\item[{14.}] D. Garcia, R. Jimenez and J. Sola, Phys. Lett. B347, 321
(1995); \\ P.H. Chankowski and S. Pokarski, Nucl. Phys. B (to be
published). 
\item[{15.}] CDF Collaboration: C. Loomis, DPF meeting, Minneapolis
(1996). 
\item[{16.}] S. Raychaudhuri and D.P. Roy, Phys. Rev. D52, 1556 (1995)
and D53, 4902 (1996).
\end{enumerate}

\newpage

\begin{enumerate}
\item[{}] Table 1: The efficiency factors for the $\ell\tau$ channel
corresponding to the indicated kinematical and topological cuts.  For
the $WW$ process, the corresponding efficiencies from the CDF
simulation are shown in parenthesis.  The middle column shows the
triggering, isolation and identification efficiencies from the CDF
simulation. 
\end{enumerate}
\medskip
\[
\begin{tabular}{|c|c|c|c|}
\hline
&&& \\
Eff. & $p^{\ell,\tau}_T$ \& geom. & $\epsilon^\ell_{tr},
\epsilon^\ell_{iso}$, & jets, $H_T$ \& $E\!\!\!/_T$ \\ 
{\rm Process} &  & $\epsilon^\ell_{id}, \
\epsilon^\tau_{id}$ &  \\
\hline
&&& \\
$WW$ & .16 (.13) & .93 $\times$ .9 & .64 (.54) \\
&&& \\
$WH (80)$ & .19 & $\times$ .87 $\times$ .5 & .61 \\
&&& \\
$WH (100)$ & .21 & $=$ .36 & .62 \\
&&& \\
$WH (120)$ & .22 & & .64 \\
&&& \\
$WH (140)$ & .22 & & .65 \\
&&& \\
\hline
\end{tabular}
\]

\vspace{5cm}

\noindent\underbar{\large{\bf Figure Captions }}\\
\smallskip
\begin{enumerate}
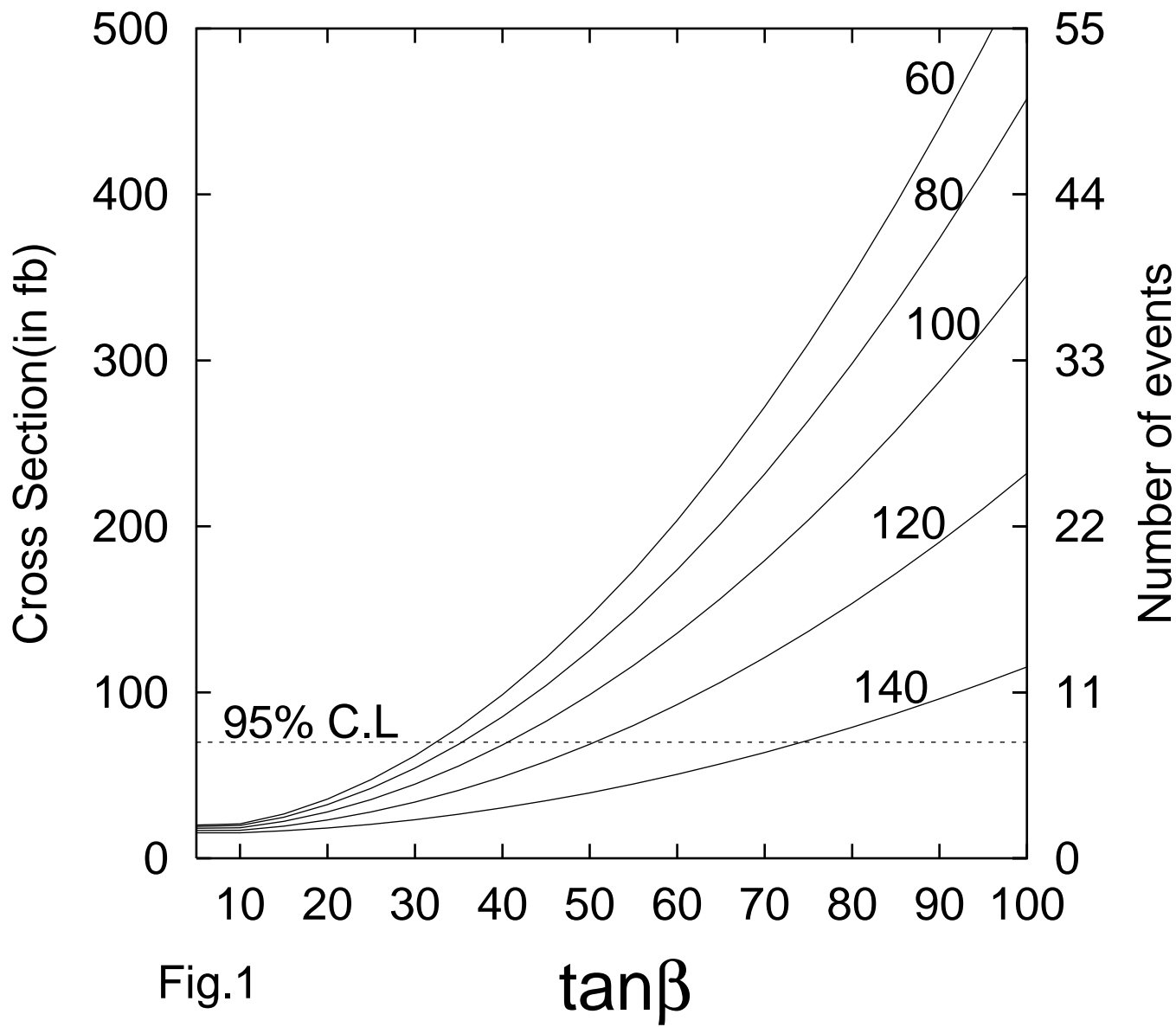
\item[{Fig.1}]:
The predicted cross-section (No. of events) shown 
against $tan\beta$ for different $H^\pm$ masses. The $95\%$ C.L limit
corresponding to 7.7 events is shown as a dashed line.
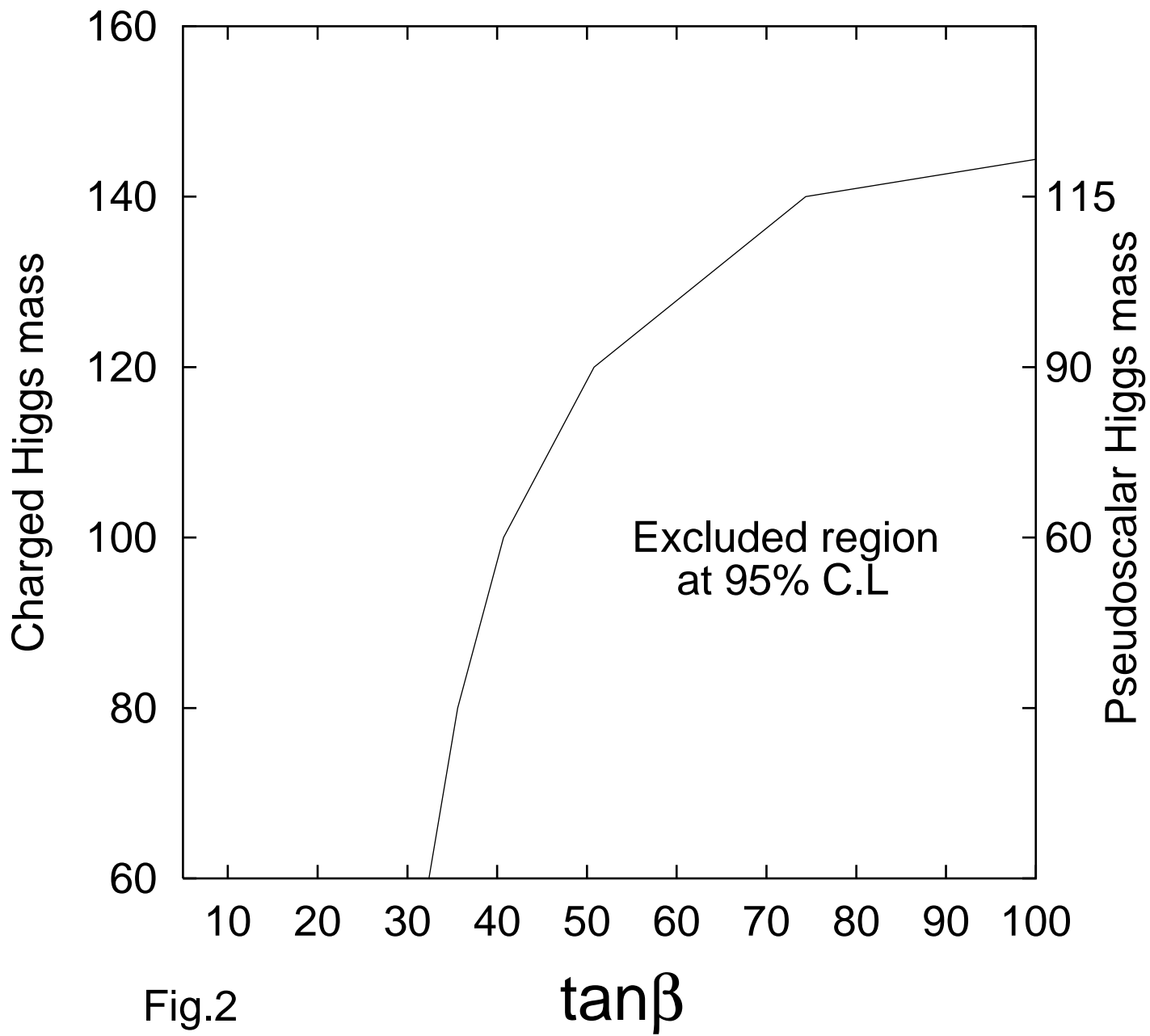
\item[{Fig.2}]:
The $95\%$ C.L exclusion contour in the $H^\pm$
mass and $tan\beta$ plane. The corresponding pseudoscalar mass 
$m_{A}$ is indicated on the right.

\end{enumerate}

\end{document}